\documentclass[12pt]{JHEP3}

\def\beq{\begin{equation}}
\def\eeq{\end{equation}}
\def\bea{\begin{eqnarray}}
\def\eea{\end{eqnarray}}
\def\nn{\nonumber}

\def\sss{\scriptscriptstyle}

\def\roughly#1{\mathrel{\raise.3ex\hbox
{$#1$\kern-.75em\lower1ex\hbox{$\sim$}}}}
\def\lsim{\roughly<}
\def\gsim{\roughly>}
\def\sla#1{\raise.15ex\hbox{$/$}\kern-.57em #1}
\def\bra#1{\left\langle #1\right|}
\def\ket#1{\left| #1\right\rangle}
\def\ks{K_{\sss S}}

\def\bd{B_d^0}
\def\bs{B_s^0}
\def\bdbar{{\bar B}_d^0}
\def\bsbar{{\bar B}_s^0}

\def\btos{{\bar b} \to {\bar s}}
\def\Bdecay{B \to J/\psi K^*}
\def\Bsdecay{\bs \to J/\psi \phi}
\def\Bpdecay{\bd \to J/\psi \ks}
\def\btopik{B \to \pi K}
\def\fT{f_{\sss T}}
\def\fL{f_{\sss L}}
\def\fTfL{f_{\sss T}/f_{\sss L}}
\def\bscc{{\bar b} \to {\bar s} c {\bar c}}

\def\gf{G_{\sss F}}
\def\lr{(\delta^d_{\sss LR})_{23}}
\def\rl{(\delta^d_{\sss RL})_{23}}

\def\bep{\varepsilon}
\def\bkll{{\bar B} \to {\bar K}^* \mu^+ \, \mu^-}
\def\AFB{A_{\rm FB}}

\preprint{UdeM-GPP-TH-09-185, UMiss-HEP-2009-01}

\title{\boldmath New Physics in $\Bsdecay$: a General
Analysis}

\author{
Cheng-Wei Chiang \\
Department of Physics and Center for Mathematics
and Theoretical Physics, \\
National Central University, Chungli, Taiwan 320, Taiwan; and \\
Institute of Physics, Academia Sinica, Taipei, Taiwan 115,
Taiwan \\
E-mail: \email{chengwei@phy.ncu.edu.tw}
}
\author{
Alakabha Datta and Murugeswaran Duraisamy \\
Department of Physics and Astronomy, 108 Lewis Hall, \\
University of Mississippi, Oxford, MS 38677-1848, USA \\
E-mail: \email{datta@phy.olemiss.edu}, \email{duraism@phy.olemiss.edu}
}
\author{
David London, Makiko Nagashima, and Alejandro Szynkman \\
Physique des Particules, Universit\'e de
Montr\'eal, C.P. 6128, succ.\ centre-ville, Montr\'eal, QC,
Canada H3C 3J7 \\
E-mail: \email{london@lps.umontreal.ca}, 
\email{makiko@lps.umontreal.ca}, 
\email{szynkman@lps.umontreal.ca}
}

\abstract{Recently, the CDF and D0 collaborations measured indirect CP
  violation in $\Bsdecay$ and found a hint of a signal. If taken at
  face value, this can be interpreted as a nonzero phase of
  $\bs$-$\bsbar$ mixing ($\beta_s$), in disagreement with the standard
  model, which predicts that $\beta_s \simeq 0$.  In this paper, we
  argue that this analysis may be incomplete.  In particular, there
  can be new physics (NP) in the $\bscc$ decay.  If so, the value of
  $\beta_s$ is different than for the case in which NP is assumed to
  be present only in the mixing.  We have examined several models of
  NP and found that, indeed, there can be significant contributions to
  the decay. These effects are consistent with measurements in
  $\Bdecay$ and $\Bpdecay$.  Due to the NP in the decay,
  polarization-dependent indirect CP asymmetries and triple-product
  asymmetries are predicted in $\Bsdecay$.}


\begin{document}

\section{Introduction}

Over the past several years, the measurements of several quantities in
a number of $B$ decays differ from the predictions of the Standard
Model (SM) by $\sim2\sigma$. For example, in $\btopik$, it is
difficult to account for all the experimental measurements within the
SM \cite{piKupdate}. Also, the SM predicts that the measured indirect
(mixing-induced) CP asymmetry in $\btos$ penguin decays should
generally be equal to that in $\bd\to J/\psi\ks$.  However, it is
found that these two quantities are not identical for several decays
\cite{LunghiSoni}. In addition, the measurement of the lepton
forward-backward asymmetry in $\bkll$ ($\AFB$) is not in perfect
agreement with the predictions of the SM \cite{AFBmeas}. Finally, in
$B\to\phi K^*$, the final-state particles are vector mesons, so that
this decay is in fact three separate decays, one for each polarization
(one longitudinal, two transverse). Naively, one expects the fraction
of transverse decays, $\fT$, to be much less than the fraction of
longitudinal decays, $\fL$. However, it is observed that these two
fractions are roughly equal: $\fTfL (B\to\phi K^{*}) \simeq 1$
\cite{phiK*}.

In most cases, the individual ``disagreements'' with the SM are not
statistically significant.  And it may be possible to explain the
value of $\fTfL$ within the SM, though this is not certain. Thus, no
real discrepancy with the SM can be claimed. Still, the differences
are intriguing since (i) there are several $B$ decays involved and a
number of different effects, and (ii) they all appear in $\btos$
transitions. (Indeed, the Belle experiment itself has claimed that the
disagreement in $\AFB$ shows a clear hint of physics beyond the SM
\cite{BellePR}.)  New-physics (NP) scenarios have been considered to
explain the possible problems. Various models have been proposed, all
of which contain new contributions to the decay ${\bar b} \to {\bar s}
q{\bar q}$ ($q=u,d$ or $s$).

Recently, a new effect has been seen in $\Bsdecay$. In the SM, the
decay of $\bd \to J/\psi \ks$ is dominated by the tree-level
transition $\bar{b} \to \bar{c} c \bar{s}$ and is real, so that
indirect CP violation probes only the phase of $\bd$-$\bdbar$ mixing,
$\beta$. This mixing phase has been measured: $\beta =
(21.58^{+0.90}_{-0.81})^\circ$ \cite{CKMfitter}. The same logic can be
applied to the $\bs$ system.  The decay of $\Bsdecay$ is similar to
that of $\bd \to J/\psi \ks$.  Thus, indirect CP violation in
$\Bsdecay$ can be used to probe the phase of $\bs$-$\bsbar$ mixing,
$\beta_s$ (which is $\simeq 0$ in the SM). The CDF \cite{CDF} and D0
\cite{D0} collaborations have presented 2-dimensional correlations of
$\beta_s$ vs.\ $\Delta\Gamma$. In a recent conference proceeding
\cite{CDFD0}, the results of the two experiments were combined, and
the CDF and D0 collaborations claimed a 2.2$\sigma$ deviation from the
prediction of the SM. This could hint at NP in $\Bsdecay$, and this is
the assumption we will use in the rest of the paper.

The phase of $\bs$-$\bsbar$ mixing therefore appears to differ from 0
by more than 2$\sigma$.  Many theoretical papers have been written
exploring the prediction for $\beta_s$ of various NP models
\cite{ZFCNC, Z'FCNC, 2HDM, SUSY, littleHiggs, unparticle, fourGen,
  Bspapers}. The main purpose of this paper is to point out that this
analysis may be incomplete.  In particular, the possibility of new
physics in the decay of $\Bsdecay$ has not been included\footnote{The
  idea of NP in the decay $\bscc$ is not new (for example, see
  Ref.~\cite{bscc}). However, its application to $\Bsdecay$ has not
  been considered previously.}.  However, this is important -- most NP
which gives a phase in $\bs$-$\bsbar$ mixing also contributes to the
decay.  As an example, consider the model with $Z$-mediated
flavor-changing neutral currents \cite{ZFCNC}.  A $Z{\bar b}s$
coupling will lead to $\bs$-$\bsbar$ mixing at tree level.  However,
this same coupling will give the (tree-level) decay $\bar{b} \to
\bar{s} c \bar{c}$, mediated by an off-shell $Z$. (In fact, in certain
models, there is little contribution to $\beta_s$; the main NP effect
is in the decay \cite{SUSY}.)  It is therefore quite natural to
consider the possibility of NP in the decay $\Bsdecay$.

However, once one does this, the conclusion regarding the phase of
$\bs$-$\bsbar$ mixing is no longer justified. While there is NP in
$\Bsdecay$ (our assumption), it need not be only in the mixing.  It
could also be in the decay, and $\beta_s$ can be different from the
case where NP contributes only to the mixing.

As we noted above, for quantities in several $B$ decays, there are
differences between the SM predictions and the central values of the
measurements, and there are NP models which account for these
differences by having new contributions to the decay ${\bar b} \to
{\bar s} q{\bar q}$ ($q=u,d, s$). In light of this, it would not be a
surprise to also find NP in $\bscc$.

Now, $\Bsdecay$ is similar to $\Bdecay$ (here, $B = \bd$ or $B^+$), as
both decays contain final-state vector mesons related by flavor SU(3)
symmetry.  This symmetry is assumed to hold approximately even in the
presence of NP. (Although the $\phi$ meson has a flavor-singlet
component, this does not cause a serious problem in relating the two
decays by SU(3).  It was shown in Ref.~\cite{Gronau:2008hb} that the
singlet diagrams in $\Bsdecay$ suffer either Cabibbo suppression or
suppression by the Okubo-Zweig-Iizuka (OZI) rule \cite{OZI}.)  Thus,
if NP contributes to $\Bsdecay$, it will also be present in $\Bdecay$,
and this will lead to new CP-violating effects (assuming that there is
at least one new weak phase).  Of course, there will be flavor SU(3)
breaking in relating the NP (and even the SM) contributions in these
decays.  However, we do not expect this effect to be very large since
the masses of the $K^*$ and $\phi$ mesons are similar: $m_{\sss K^*} =
892$ MeV, $m_\phi = 1020$ MeV.

In the decay $\Bdecay$, CP-violating effects have been looked for
experimentally, but none has been found.  This can put constraints on
any NP contribution to the decay in $\Bsdecay$.  There are two
possible conclusions for a given model.  First, the constraints might
be so strong that any significant contribution of the model to the
decay is ruled out.  Second, the constraints might be
weak enough that a NP contribution to the decay of $\Bsdecay$ is still
possible.  In this case, if the contribution is sizeable, the CDF/D0
analysis will have to take into account NP in both the mixing and the
decay. Alternatively, if this contribution is small, the CDF/D0
analysis applies directly. In both cases, it is important to ascertain
if there are any other predictions of the model. {}From this, we see
that it is necessary to examine all NP models to see which of these
possibilities occurs.

In Sec.~\ref{sec2}, we present the results of the experimental
searches for CP violation in $\Bdecay$.  The absence of evidence for
such CP violation can lead to constraints on NP in the decay of
$\Bsdecay$. We also discuss possible future measurements. These can be
used to confirm the presence of NP and to distinguish different
models. In Sec.~\ref{sec3} we look at several models which lead to NP
in the mixing and/or the decay of $\Bsdecay$. We examine the effect of
the constraints from $\Bdecay$, and look at predictions of further
effects.  We conclude in Sec.~\ref{sec4}.

\section{\boldmath CP Violation in $\Bdecay$ \label{sec2}}

The decay $\Bdecay$ is really three separate decays, one for each
polarization state $\lambda$ of the final-state vector particles;
longitudinal: $\lambda = 0$, transverse: $\lambda = \left\{\|,\perp
\right\}$.

Suppose now that there are several new-physics amplitudes, each with a
different weak phase, that contribute to the decay. In
Ref.~\cite{DLNP}, it is argued that all strong phases associcated with
NP amplitudes are negligible.  The reason is that the strong phases
are generated by rescattering, and this costs a factor of about
25. The strong phase of the SM color-suppressed ${\bar b} \to {\bar c}
c {\bar s}$ diagram $C$ is generated by rescattering of the
color-allowed ${\bar b} \to {\bar c} c {\bar s}$ tree diagram $T$.
Since $|C/T|$ is expected to be in the range 0.2-0.6, the SM strong
phase is on the small side, but is not negligible.  On the other hand,
the strong phases of NP amplitudes can only be generated by
rescattering of the NP diagrams themselves ({\it i.e.},
self-rescattering) at low energies.  They are therefore very small,
and can be neglected. In this case, for each polarization one can
combine all NP matrix elements into a single NP amplitude, with a
single weak phase $\varphi_\lambda$:
\beq
\sum \bra{(J/\psi K^*)_\lambda} {\cal O}_{\sss NP} \ket{B}
= b_\lambda e^{i\varphi_\lambda} ~.
\label{ONP}
\eeq
It must be emphasized that the above argument for negligible strong
rescattering phases associated with NP amplitudes complies with the
operator product expansion formalism, and does not rely on
factorization.  Moreover, the validity of the argument can be checked
experimentally by carefully measuring and comparing direct CP
asymmetries and triple-product correlations in $\Bsdecay$ and
$\Bdecay$, because of their different dependences on the weak and
strong phases.

We now assume that this single NP amplitude contributes to the decay.
The decay amplitude for each of the three possible polarization states
may then be written as
\bea
A_\lambda \equiv Amp (\Bdecay)_\lambda &=&
a_\lambda e^{i (\delta_\lambda^a -
\delta_\perp^a)} + b_\lambda e^{i\varphi_\lambda} e^{- i
\delta_\perp^a} ~, \nn\\
{\bar A}_\lambda \equiv Amp ({\bar B} \to J/\psi {\bar
K}^*)_\lambda &=&
a_\lambda e^{i (\delta_\lambda^a -
\delta_\perp^a)} + b_\lambda e^{-i\varphi_\lambda} e^{- i
\delta_\perp^a} ~,
\label{amps}
\eea
where $a_\lambda$ and $b_\lambda$ represent the SM and NP amplitudes,
respectively, $\varphi_\lambda$ is the new-physics weak phase, and the
$\delta_\lambda^a$ are the SM strong phases.  All strong phases are
given relative to $\delta_\perp^a$. $a_\lambda$ is defined to be
positive for every polarization. $b_\lambda$ can also be taken to be
positive: if it is negative, the minus sign can be absorbed in the
weak phase by redefining $\varphi_\lambda \to \varphi_\lambda +
\pi$. We emphasize this fact by writing the ratio
$b_\lambda/a_\lambda$ as the positive-definite quantity $|r_\lambda|$.

Note: if there is only one NP operator, the weak phase can be taken to
be polarization-independent. We write it simply as $\varphi$. (Several
of the models studied in the next section are of this type.)  However,
one has to be careful here: in this case the $b_\lambda$'s cannot be
taken to be positive since, if one $b_\lambda$ is negative, the minus
sign cannot be removed by redefining $\varphi$.

The polarization amplitudes can be extracted experimentally by
performing an angular analysis of $\Bdecay$, in which the $J/\psi$ and
$K^*$ are detected through their decays to $\ell^+\ell^-$ and $P_1
{\bar P}_2$, respectively (the $P_i$ are pseudoscalars) \cite{DDF}. We
take the $K^*$ linear polarization vector to lie in the $x$-$y$ plane.
We then define the angle $\psi$ to be that of the $P_1$ in the $K^*$
rest frame relative to the polarization axis (the negative of the
direction of the $J/\psi$ in that frame). The $K^*$ has a single
linear polarization state $\bep$ for each amplitude: in the $J/\psi$
rest frame,
\beq
A_\|: ~~\bep = \hat y~~~;~~~
A_0        : ~~\bep = \hat x~~~;~~~
A_\perp    : ~~\bep = \hat z~.
\eeq
A unit vector ${\hat n}$ in the direction of the $\ell^+$ in $J/\psi$
decay is defined to have components
\beq \label{eqn:ndef}
(n_x, n_y, n_z) = (\sin \theta \cos \varphi,
\sin \theta \sin \varphi, \cos \theta)~,
\eeq
where $\varphi$ is the angle between the projection of the $\ell^+$ on
the $P_1 {\bar P}_2$ plane in the $J/\psi$ rest frame and the $x$
axis.

The angular distribution is then
\bea
\label{angdist}
\frac{d^4 \Gamma [B \to (\ell^+\ell^-)_{\sss J/\psi} (P_1 {\bar
P}_2)_{\sss K^*}]} {d \cos \theta~d \varphi~d \cos \psi~dt}
& = &  \frac{9}{32 \pi} \left[ 2 |A_0|^2 \cos^2 \psi
  (1 - \sin^2 \theta \cos^2 \varphi) \right. \nn\\
&& \hskip-4truecm + ~ \sin^2 \psi \{ |A_\parallel|^2 (1 -
\sin^2 \theta \sin^2 \varphi) + |A_\perp|^2 \sin^2 \theta
- {\rm Im}(A_\parallel^* A_\perp) \sin 2
\theta \sin \varphi \} \nn\\
&& \hskip-4truecm \left. +~\frac{1}{\sqrt{2}} \sin 2 \psi \{ {\rm
Re}(A_0^* A_\parallel) \sin^2 \theta \sin 2 \varphi
+{\rm Im}(A_0^* A_\perp) \sin 2 \theta
\cos \varphi \} \right]~.
\eea
For $\bar B$ decays, the interference terms involving the $A_\perp$
amplitude are of opposite sign and all other terms are unchanged.  The
angular distribution is the same for $\Bsdecay$, where the $\phi$ is
detected via its decay to $K^+ K^-$.

In order to probe the NP amplitudes $b_\lambda$ of Eq.~(\ref{amps}),
one has to measure CP-violating observables.  The most obvious of
these is the direct CP asymmetry, in which one compares the rates for
process and anti-process.  Ideally, this would be measured for each of
the three polarization states individually.  Unfortunately, this has
not been done at BaBar or Belle (although they have both measured the
angular distribution). BaBar has measured the direct CP asymmetry for
the entire process, combining $\lambda = 0, \|, \perp$.  They find a
result of $(2.5 \pm 8.3 \pm 5.4)\%$ \cite{BaBarDCPV}, which is
consistent with zero.  This could suggest that each of the
$b_\lambda$'s is tiny, {\it i.e.}, that there is essentially no NP in
$\bscc$.  However, the direct CP asymmetry is proportional to the sine
of the difference of the strong phases of the interfering amplitudes,
{\it i.e.}, $\sin \delta_\lambda^a$.  As we have argued above, this is
expected to be small.  As a result, the direct CP asymmetry will also
be small, not because the NP amplitudes are absent, but rather due to
the size of the strong phases.  Thus, the direct CP asymmetry cannot
be used to constrain the $b_\lambda$'s.

It should be noted that BaBar finds a rather large strong-phase
difference: $\delta_\|^a - \delta_\perp^a = (25.8 \pm 2.9 \pm
1.1)^\circ$ \cite{BaBarang}.  This is confirmed by Belle
\cite{Belleang}.  However, this is misleading. BaBar has implicitly
assumed that there is no NP in the decay -- this is indicated by the
presentation of results as a ``strong-phase difference.''  What really
is measured is the difference ${\rm Arg}(A_\|) - {\rm Arg}(A_\perp)$
(which is how Belle presents its results), and this could have a
contribution from NP.  In addition, because the SM is assumed by both
collaborations, the data from the decay and charge-conjugate decay are
added together.  But in the presence of NP with a new weak phase,
these are different.  The upshot is that this measurement does not
take into account the possibility of NP in the decay, and so is not
relevant to our analysis.

Now, even though the direct CP asymmetry is not used here, there is
fortunately another CP-violating observable which is pertinent.  It
involves the triple-product correlation (TP) \cite{DattaLondon}. In
the rest frame of the $B$, the TP takes the form ${\vec q} \cdot
({\vec\bep}_{\sss J/\psi} \times {\vec\bep}_{\sss K^*})$, where ${\vec
  q}$ is the momentum of one of the final vector mesons, and
${\vec\bep}_{\sss J/\psi}$ and ${\vec\bep}_{\sss K^*}$ are the
polarizations of the $J/\psi$ and $K^*$.

TP asymmetries are similar to direct CP asymmetries in that both are
obtained by comparing a signal in the process with the corresponding
signal in the anti-process, and both are nonzero only if there are two
interfering decay amplitudes. However, whereas the direct CP asymmetry
is ${\cal A}_{\sss CP}^{dir}|_\lambda \propto \sin\varphi_\lambda
\sin\delta_\lambda^a$, the TP asymmetry ${\cal A}_{\sss TP}$ involves
the product $\sin\varphi_{\lambda'} \cos\delta_{\lambda}^a$.  That is,
while the direct CP asymmetry is produced only if there is a nonzero
strong-phase difference between the two decay amplitudes, the TP
asymmetry is {\it maximal} when the strong-phase difference
vanishes. Thus, given that the strong phases are expected to be small,
TP asymmetries are particularly promising for $B$ decays.

There are two TPs in $\Bdecay$; they are proportional to ${\rm
  Im}(A_\perp A_0^*)$ and ${\rm Im}(A_\perp A_\|^*)$
\cite{DattaLondon}.  These two terms appear in the angular
distribution [Eq.~(\ref{angdist})], and can thus be obtained through
this measurement.  The TPs are defined as
\bea
A_{\sss T}^{(1)} & \equiv & \frac{{\rm Im}(A_\perp
A_0^*)}{|A_0|^2+|A_\| |^2+|A_\perp|^2} ~, \nn\\
\label{TPmeasure}
A_{\sss T}^{(2)} & \equiv & \frac{{\rm Im}(A_\perp
A_\|^*)}{|A_0|^2+|A_\| |^2+|A_\perp|^2} ~.
\label{TPs}
\eea 
The corresponding quantities for the charge-conjugate process,
$\bar{A}_{\sss T}^{(1)}$ and $\bar{A}_{\sss T}^{(2)}$, are defined
similarly.  The TP asymmetries are obtained by calculating the
difference of each of the above TPs and that in the anti-process.

As noted above, both BaBar and Belle have measured the angular
distribution.  As such, they have obtained the TP asymmetries for a
variety of $K^*$ decays \cite{BaBarang,Belleang}.  Averaging the
results, they find
\bea
{\cal A}_{\sss TP}^{(1)} & = & 0.017 \pm 0.033 ~~(\bd) ~,
\nn\\
& = & 0.013 \pm 0.053 ~~(B^+) ~, \nn\\
{\cal A}_{\sss TP}^{(2)} & = & -0.004 \pm 0.025 ~~(\bd) ~,
\nn\\
& = & -0.014 \pm 0.030 ~~(B^+) ~.
\label{TPresults}
\eea
All the measured TP asymmetries are consistent with zero. In contrast
to the direct CP asymmetry, these measurements cannot be explained by
small strong phases.

In the introduction, we noted that any NP in the decay of $\Bsdecay$
would also be present in $\Bdecay$.  Thus, results in $\Bdecay$ can be
used to constrain the NP. We conservatively incorporate the
measurements of Eq.~(\ref{TPresults}) by requiring that the
predictions of TPs in $\Bsdecay$ obey $|{\cal A}_{\sss TP}^{(1,2)}|
\le 10\%$ (this value takes into account the errors in
Eq.~(\ref{TPresults}), as well as a possible SU(3)-breaking effect in
relating the two decays).

Above, we have argued that the strong phases are expected to be small,
and will therefore not produce a small TP asymmetry since this is
proportional to $\cos\delta_\lambda^a$.  On the other hand, one might
wonder whether nonperturbative effects might play a role, and lead to
$\delta_\lambda^a \simeq \pi/2$.  If so, they would lead to ${\cal
  A}_{\sss TP}^{(i)} \simeq 0$. However, one can also produce a ``fake
TP asymmetry,'' which is calculated from the {\it sum} of each of the
TPs in Eq.~(\ref{TPmeasure}) and that in the anti-process. This
quantity ${\cal A}_{\sss TP}^{fake}$ involves the product
$\cos\varphi_{\lambda'} \sin\delta_{\lambda}^a$, and has been measured
by Belle \cite{Belleang}:
\bea
{\cal A}_{\sss TP}^{fake, (1)} & = & 0.138 \pm 0.046 ~~(\bd)
~, \nn\\
& = & 0.108 \pm 0.054 ~~(B^+) ~, \nn\\
{\cal A}_{\sss TP}^{fake, (2)} & = & -0.187 \pm 0.043 ~~(\bd)
~, \nn\\
& = & -0.080 \pm 0.052 ~~(B^+) ~.
\eea
The fake TP asymmetries are nonzero, pointing to a nonzero value of
the strong phase $\delta_\lambda^a$.  But these are all fairly small,
so that $\delta_\lambda^a \simeq \pi/2$ is not allowed.  This rules
out the possibility of nonperturbative effects leading to large strong
phases.

Finally, a full time-dependent angular analysis of $\bd \to J/\psi
K^{*0} ~~ (K^{*0} \to \ks \pi^0)$ has not yet been done.  However, if
it can be performed, there are many more tests for NP in the decay.
This is discussed in detail in Ref.~\cite{LSS}; we summarize the
results briefly below.

The time-dependent decay rates can be written as
\bea
\Gamma(\bd(t) \to J/\psi K^{*0}) & \!=\! & e^{-\Gamma t}
\sum_{\lambda\leq\sigma} \Bigl(\Lambda_{\lambda\sigma} +
\Sigma_{\lambda\sigma}\cos(\Delta M t) -
\rho_{\lambda\sigma}\sin(\Delta M t)\Bigr) g_\lambda g_\sigma
~,~~ \nn\\
\Gamma(\bdbar(t) \to J/\psi {\bar K}^{*0}) & \!=\! & e^{-\Gamma t}
\sum_{\lambda\leq\sigma} \Bigl(\Lambda_{\lambda\sigma} -
\Sigma_{\lambda\sigma}\cos(\Delta M t) +
\rho_{\lambda\sigma}\sin(\Delta M t)\Bigr) g_\lambda g_\sigma
~,
\label{decayrates}
\eea
where the $g_\lambda$ are the coefficients of the helicity amplitudes
written in the linear polarization basis. The $g_\lambda$ depend only
on the angles describing the kinematics \cite{glambda}. By performing
the above time-dependent angular analyses, one can measure the 18
observables $\Lambda_{\lambda\sigma}$, $\Sigma_{\lambda\sigma}$ and
$\rho_{\lambda\sigma}$ ($\lambda\leq\sigma$). Not all of these are
independent.  There are a total of six amplitudes describing $\bd,
\bdbar \to J/\psi K^{*0}$ decays. At best one can measure the
magnitudes and relative phases of these six amplitudes, giving 11
independent measurements.

In the absence of NP, the $b_\lambda$ are zero in Eq.~(\ref{amps}).
The number of parameters is then reduced from 13 to 6: three
$a_\lambda$'s, two strong-phase differences, and the weak phase of
$\bd$-$\bdbar$ mixing, $\beta$. Given that there are 18 observables,
there must exist 12 relations among the observables in the absence of
NP. These are:
\bea
&& \Sigma_{\lambda\lambda} = \Sigma_{\| 0}=0 ~,~~~~
\Lambda_{\perp i} = 0 ~,\nn\\
&& \frac{\rho_{ii}}{\Lambda_{ii}} =
-\frac{\rho_{\perp\perp}}{\Lambda_{\perp\perp}} =
\frac{\rho_{\|0}}{\Lambda_{\| 0}} ~, \nn \\
&& \Lambda_{\|0}=\frac{1}{2\Lambda_{\perp\perp}}\Bigl[
\frac{\Lambda_{\lambda\lambda}^2\rho_{\perp 0}
\rho_{\perp\|}+\Sigma_{\perp 0}\Sigma_{\perp
\|}(\Lambda_{\lambda\lambda}^2 -\rho_{\lambda\lambda}^2)}
{\Lambda_{\lambda\lambda}^2-\rho_{\lambda\lambda}^2}\Bigr] ~, \nn \\
&& \frac{\rho_{\perp
i}^2}{4\Lambda_{\perp\perp}\Lambda_{ii}-\Sigma_{\perp
i}^2}=\frac{\Lambda_{\perp\perp}^2
-\rho_{\perp\perp}^2}{\Lambda_{\perp\perp}^2} ~,
\label{eq:no_np}
\eea
where $i=\{0,\|\}$. The violation of any of the above relations is a
smoking-gun signal of NP in the decay.

The first line in Eq.~(\ref{eq:no_np}) simply states that, if NP is
absent, the direct CP asymmetries ($\Sigma_{\lambda\lambda}$,
$\Sigma_{\| 0}$) and the triple-product asymmetries ($\Lambda_{\perp
  i}$) all vanish.  This has been discussed earlier.  It is the second
line which is particularly interesting.  It states that the indirect
CP asymmetry for each polarization state should be the same within the
SM (all giving $\sin 2\beta$ in $\Bdecay$, modulo an overall sign for
the $\perp$ polarization).  If a different result for different
polarizations is found, this would be a clear sign of NP in the decay.
Although this measurement has not yet been made, it might be done in
the future at LHCb.

As we will see in the next section, some models of NP in $\Bsdecay$ do
predict polarization-dependent indirect CP asymmetries in both
$\Bdecay$ and $\Bsdecay$. This property of the NP should be taken into
account when analyzing the $\Bsdecay$ data. In analyzing the CDF/D0
data, it is important {\it not} to average over polarizations. Thus,
not only should any analysis take into account the possibility of NP
in the decay, it should also consider the case where the NP is
polarization-dependent.

Note: in Eq.~(\ref{decayrates}) and elsewhere throughout the paper, we
neglect the width difference in the $\bs$ system, $\Delta\Gamma_s$.
Now, there are theoretical estimates of $\Delta\Gamma_s$ in the SM
\cite{DeltaGammasth1,DeltaGammasth2}. For example, in
Ref.~\cite{DeltaGammasth1} it is found that, for a typical set of
parameters, $\Delta\Gamma_s/\Gamma_s = 0.15 \pm 0.06$.  The central
value is significant, suggesting that $\Delta\Gamma_s$ might not be
negligible. However, the theoretical error is also large, so that the
prediction of a sizeable width difference is not certain. In
principle, this uncertainty can be resolved by an experimental
measurement. Unfortunately, here too the errors are very large
\cite{DeltaGammasexp}. Thus, at present, there is no clear
experimental or theoretical result suggesting a large value of
$\Delta\Gamma_s$, so that our neglect of this quantity is justified.
However, should this change in the future, for example through a
direct measurement at LHCb, it will be necessary to take
$\Delta\Gamma_s$ into account in the calculations.

\section{\boldmath Models of New Physics in $\Bsdecay$ \label{sec3}}

In this section, we examine several models of NP which have been
proposed to produce large $\bs$-$\bsbar$ mixing.  In all cases, the
aim is to see whether significant contributions to the decay $\bscc$
are also possible, consistent with constraints from $\Bdecay$.

In order to determine whether the contribution to the decay is
``significant,'' we need to compare its apparent effect on
$\bs$-$\bsbar$ mixing (as deduced from the indirect CP asymmetry) with
the best-fit measured value, $\beta_s^{meas}$. Unfortunately, though
the CDF and D0 collaborations noted that there is a 2.2$\sigma$ effect
in indirect CP violation in $\Bsdecay$ \cite{CDFD0}, they never gave a
central value for $\beta_s^{meas}$.  There is, however, an
alternative. The UTfit Collaboration \cite{UTfit} analyzed the CDF/D0
data, and found favored values for $\beta_s^{meas}$. It should be
noted that the UTfit analysis is not universally accepted. They found
an effect larger than 3$\sigma$, which is at odds with that found by
the experimental collaborations themselves. (And, in obtaining
$\beta_s^{meas}$, the UTfit group averaged over polarizations, which,
as explained above, is not completely general.) Still, we need only a
preferred value for $\beta_s^{meas}$, not the error.  For this reason,
in our numerical analysis below, we use the UTfit best-fit value.  But
the analysis can be straightforwardly repeated for any other value of
$\beta_s^{meas}$. The UTfit analysis finds that $\bs$-$\bsbar$ mixing
obeys $\sin2\beta_s^{meas} = -(0.6 \pm 0.2)$ [S1: $\beta_s^{meas} =
  (-19.9 \pm 5.6)^\circ$] or $-(0.7 \pm 0.2)$ [S2: $\beta_s^{meas} =
  (-68.2 \pm 4.9)^\circ$].

If there is NP in the decay, it will contribute to
the indirect CP asymmetry. In the presence of a nonzero $b_\lambda$
[Eq.~\ref{amps}], the general expression for the result of the
indirect CP asymmetry in $\Bsdecay$ is
\bea
\sin {2\beta}_s^{meas} & = &
\frac{\strut \sin2\beta_s + 2
|r_\lambda| \cos\delta_{\lambda}^a
\sin(2\beta_s+\varphi_\lambda)+ |r_\lambda|^2
\sin(2\beta_s+\varphi_\lambda) } {\strut 1 + 2 |r_\lambda|
\cos\delta_{\lambda}^a \cos\varphi_\lambda + |r_\lambda|^2 },
\eea
where $|r_\lambda| \equiv b_\lambda/a_\lambda$ (although we retain the
index $\lambda$, in the following we ignore the polarization
dependence). (There is an additional overall minus sign for the
$\perp$ polarization.)

For small $|r_\lambda|$ [neglecting $O(|r_\lambda|^2)$], we have
\bea
\sin{2 \beta}_s^{meas} & = & [\sin2\beta_s + 2 |r_\lambda|
 \cos\delta_{\lambda}^a \sin(2\beta_s+\varphi_\lambda)]
[1 - 2 |r_\lambda| \cos\delta_{\lambda}^a
 \cos\varphi_\lambda] \nn\\
& = & \sin2\beta_s + 2 |r_\lambda| \cos{2 \beta_s}
 \sin\varphi_\lambda\cos\delta_{\lambda}^a ~.
\label{beta-lambda}
\eea
In the SM, $\beta_s \simeq 0$.  Taking $\delta_{\lambda}^a \simeq 0$,
we see that NP in the decay can reproduce the observed value of
$\sin{2 \beta}_s^{meas}$ if $|r_\lambda| \simeq 30\%$ (and
$\sin\varphi_\lambda \simeq -1$).  However, even if $\bs$-$\bsbar$
mixing arises from NP contributions (so that $\beta_s \to
\beta_s^{\sss NP} \ne 0$ and $\cos{2 \beta_s^{\sss NP}} < 1$), the
contribution from the decay is still a non-negligible fraction of
$\sin{2 \beta}_s^{meas} = 0.6$-$0.7$ if $|r_\lambda| \gsim 10\%$.  If
it is found that $|r_\lambda|$ satisfies this limit, we consider the
contribution to the decay significant.

Given that there is NP in the decay $\bscc$, it is important to check
that the measurement of $\beta$ in $\Bpdecay$ is still consistent. In
the presence of NP in the decay ($|r|\ne 0$), the effective measured
$\sin {2 \beta}$ in $\Bpdecay$ is given by
\beq
\sin{2 \beta}^{meas} = \sin2\beta + 2 |r| \cos{ 2 \beta}
\sin\varphi\cos\delta^a ~.
\label{s}
\eeq
The (true) value of $\sin {2 \beta}$ can be taken from the fit to the
sides of the unitarity triangle: $\sin {2 \beta}=0.731 \pm 0.038$,
while the experimental measurement gives $\sin{2 \beta}^{meas}=0.668
\pm 0.028$ \cite{CKMfitter}.  Taking $\delta^a \simeq 0$ and
$\sin\varphi \simeq -1$, we obtain
\bea
|r| & = & \frac{\sin{2 \beta}^{meas} - \sin2\beta}{-2 \cos{ 2
    \beta}}
= (4.6 \pm 3.5) \% ~.
\label{ratio}
\eea
Allowing up to a $3\sigma$ variation, we see that $|r| \le 15\%$ is
permitted.

In a recent article \cite{Soni2009}, it was suggested that, based on
lattice calculations, the true value of $\sin {2 \beta}$ is even
larger, perhaps up to 0.87.  In this case, even bigger values of $|r|$
are allowed.

The point here is that some authors have claimed that there is an
apparent discrepancy between the measured value of $\sin {2 \beta}$
and the true, underlying value, and that this calls for NP in
$\bd$-$\bdbar$ mixing.  What we have seen above is that this
discrepancy might be due to NP in the $\bscc$ decay.

For all models, we will compare their contributions to the decay to
those of the SM. The SM operator is $LL$ and is given by
\beq
\frac{\gf}{\sqrt{2}}
\left(c_2 + \frac{c_1}{N_c} \right) V_{cb}^* V_{cs} \, \bar s
\gamma_\mu (1-\gamma_5) b \, \bar c \gamma^\mu (1-\gamma_5) c
~,
\label{Xsm}
\eeq
where $c_{1,2}$ are Wilson coefficients.  We have $c_1(m_b) = 1.081$,
$c_2(m_b) = -0.190$, so that $c_2 + c_1/N_c = 0.17$.  Also, $|V_{cb}^*
V_{cs}| = 0.041$.  Thus, the coefficient of the SM amplitude is
$0.007$ (here and below, we ignore the factor $\gf/\sqrt{2}$).

Within factorization, the SM matrix elements are given by $\langle K^*
| \bar s \gamma_\mu (1-\gamma_5) b | B \rangle$ $\langle J/\psi | \bar
c \gamma^\mu (1-\gamma_5) c | 0 \rangle$. Since the $J/\psi$ is a
vector meson, $\langle J/\psi | \bar c \gamma^\mu \gamma_5 c | 0
\rangle = 0$, {\it i.e.}, the axial-vector piece vanishes.  Thus, the
SM operator is really $LV$. The calculation of $|r_\lambda|$ includes
the ratio of NP and SM matrix elements.

Now, throughout the paper, we perform the calculations in the context
of factorization.  But there may be some concerns about the use of
factorization in decays in which $J/\psi$ mesons are involved.  For
example, the decay $\Bdecay$ was studied in QCD factorization in
Ref.~\cite{cheng}.  It was found that naive factorization is unable to
explain the branching ratio and the various polarization fractions in
this decay.  In addition, the quantity $a_2 = c_2+ c_1/N_c$ can be
extracted from experiment \cite{cheng}, but it is found to be
dependent on the polarization. This could lead to errors in the
predictions of various NP models for the polarization-dependent
indirect CP asymmetries. However, in all cases, the effects are small.
Still, the results of our studies should be understood as estimates
rather than precise calculations.

Finally, if there is NP in the decay $\bscc$, one might be concerned
about its effect on the width difference $\Delta \Gamma_s$ in the
$\bs$ system. If this were significant, the NP could be detected by
the measurement of this difference. Fortunately, the NP discussed in
this paper does {\it not} contribute significantly to $\Delta
\Gamma_s$.  The NP contribution to $\bscc$ is at most 15\% that of the
SM. But the SM contribution in this case comes from the
color-suppressed diagram, $C$. On the other hand, the main
contribution to $\Delta \Gamma_s$ is due to the color-allowed diagram,
$T$. Since $|C/T|$ is about 20\%, the NP contribution to $\Delta
\Gamma_s$ is only at the percent level, and is negligible.

\subsection{\boldmath $Z$-mediated FCNC's}

In the model with $Z$-mediated FCNC's ($Z$FCNC), it is assumed that a
new vector-like iso\-singlet down-type quark $d'$ is present.  Such
quarks appear in $E_6$ GUT theories, for example. The ordinary quarks
mix with the $d'$.  As a result, FCNC's appear at tree level in the
left-handed sector.  In particular, a $Z{\bar b}s$ coupling can be
generated:
\beq
{\cal L}^{\sss Z}_{\sss FCNC} = -\frac{g}{2 \cos\theta_{\sss
W}} U_{sb} \, \bar s_{\sss L} \gamma_\mu b_{\sss L} \, Z^\mu
+ {\rm h.c.}  
\eeq
This coupling leads to a NP contribution to $\bs$-$\bsbar$ mixing at
tree level.  In Ref.~\cite{ZFCNC}, it is found that one can reproduce
the measured value of $\Delta M_s$ if $|U_{sb}| \simeq 0.002$.

This coupling will also lead to the decay $\bscc$ at tree level,
mediated by a virtual $Z$.  The amplitude is
\beq
\frac{\gf}{\sqrt{2}} \, U_{sb} \, \bar s \gamma_\mu (1-\gamma_5)
b \, \bar c \gamma^\mu (I_3 - Q \sin^2\theta_{\sss W}) (1\mp
\gamma_5) c ~.
\eeq
There are thus two types of NP operators, $O_{\sss LL}$ and $O_{\sss
  LR}$, depending on whether the $c$ quark is left- or right-handed.

However, above we have noted that the matrix element $\langle J/\psi |
\bar c \gamma^\mu \gamma_5 c | 0 \rangle$ vanishes, so that it is only
the $\gamma^\mu$ piece of both of these operators which contributes.
In other words, both operators are proportional to $O_{\sss LV}$ (as
in the SM), and they can therefore be combined. In addition, as we
have noted previously, since there is only one operator, the NP weak
phase $\varphi$ can be taken to be polarization-independent.

In summary, within factorization, the total Hamiltonian can be written
as
\bea
H_{eff}^{tot} & = & H_{eff}^{\sss SM}+H_{eff}^{\sss NP,Z}
~,\nn\\
H_{eff}^{\sss SM} &= & \frac{\gf}{\sqrt{2}} \, V_{cb}V_{cs}^*
\, a_2 \, V_{\sss LV} + {\rm h.c.}, \nn\\
H_{eff}^{\sss NP,Z} & = & \frac{\gf}{\sqrt{2}} \,
V_{cb}V_{cs}^* \, a_2 \, a_{\sss Z} \, V_{\sss LV} + {\rm
h.c.}, \nn\\
V_{\sss LV} & = & \bar{s}\gamma_{\mu} ( 1-\gamma_5) b \,
\bar{c}\gamma^{\mu} c ~,
\label{heff_final_Z}
\eea
where $a_2 = c_2 + c_1/N_c = 0.17$, and
\beq
\left \vert a_{\sss Z} \right \vert = \left \vert \left(
\frac{1}{2} -\frac{4}{3} \sin^2\theta_{\sss W}
\right)\frac{U_{sb}}{V_{cb}V_{cs}^* \, a_2} \right\vert
\simeq 0.06 ~.
\label{aZ} 
\eeq
Thus, compared to the SM, the contribution of the $Z$FCNC model is
about 6\%.  This is not very large. We therefore see that the model
with $Z$-mediated FCNC's does not lead to significant new effects in
the decay.

Since this contribution is proportional to $V_{\sss LV}$, as in the
SM, the ratio $b_\lambda/a_\lambda$ is in fact {\it independent} of
$\lambda$; we simply refer to it as $r$. Furthermore, the NP and SM
matrix elements cancel in this ratio.

We can now calculate various quantities with the above effective
Hamiltonian. However, before doing so, it is useful to re-examine the
amplitudes in some detail.  Within factorization, there is no
rescattering, and the strong phases are zero.  Since the NP
contribution has the same form as that of the SM, there is a relative
$+$ sign between the two.  Also, due to form factors, etc., the
amplitude $A_0$ has an additional $-$ sign \cite{cheng}.  When one
includes nonfactorizable effects, the SM strong phases become nonzero.
Dropping the global phase, the polarization amplitudes then take the
form
\bea
A_0 & = & a_0 [ e^{i \delta_0^a}
- |r| e^{i\varphi} ] ~, \nn\\
A_\perp & = & a_\perp [ e^{i
\delta_\perp^a} + |r| e^{i\varphi} ] ~, \nn\\
A_\| & = & a_\| [ e^{i
\delta_\|^a} + |r| e^{i\varphi} ] ~,
\eea
where $\delta_0^a$ is near $\pi$, and $\delta_\perp^a$ and
$\delta_\|^a$ are near 0.

The experiments have measured differences of phases and give the
results in terms of strong-phase differences: $\delta_\perp^a -
\delta_0^a \simeq 180^\circ$, $\delta_\|^a - \delta_\perp^a \simeq
21^\circ$ in $\Bsdecay$ \cite{CDF,D0}, with similar results for
$\Bdecay$ \cite{BaBarang,Belleang}.  As discussed earlier, what is
really measured is the total phase difference between {\it
  amplitudes}, and this could, in principle, have a contribution from
NP.  However, because $|r|$ is small, the effect of NP is also small,
and so the phase differences between total amplitudes are
approximately equal to the strong-phase differences between the SM
amplitudes $\delta_\lambda^a$.  Thus, the experimental measurements
confirm the approximate factorization results that $\delta_0^a \simeq
\pi$, $\delta_\perp^a, \delta_\|^a \simeq 0$.

We now turn to the triple products.  As noted earlier, both the SM and
NP contributions are proportional to $O_{\sss LV}$.  Thus, in the
factorization limit (no strong phases), the interfering amplitudes are
kinematically identical, and so the TPs in $\Bdecay$ vanish.  When the
SM strong phases are included, one can in fact generate TPs.  However,
the TP asymmetries are proportional to ($\cos\delta_\perp^a \mp
\cos\delta_i^a$) [the $-$ ($+$) sign applies to $i = \|$ ($i = 0$)].
While this is now nonzero, it is small.  Thus, the TP asymmetries in
the $Z$FCNC model are small, and there are no constraints from the
measurements of TP asymmetries in $\Bdecay$.

Since there is a contribution to the $\bscc$ decay, there could in
principle be a polarization-dependent indirect CP asymmetry in
$\Bsdecay$ or $\Bdecay$.  However, as noted above, the ratio
$|r_\lambda|$ and the weak phase $\varphi$ are in fact
polarization-independent, so that no such effect is predicted.

This is confirmed by explicit calculation.  As shown previously, in
the presence of NP in the decay, the indirect CP asymmetry in
$\Bsdecay$ is given by
\beq
\sin{2 \beta}_s^{meas} = \sin2\beta_s^{\sss NP} \pm 2 |r| 
\cos{2 \beta_s^{\sss NP}}
 \sin\varphi\cos\delta_{\lambda}^a ~,
\label{sin2betaZ}
\eeq
where the $+$ ($-$) sign applies to $\lambda = \perp,\|$ ($\lambda =
0$). Taking $\delta_0^a \simeq \pi$, $\delta_\perp^a, \delta_\|^a
\simeq 0$, we see that the correction to $\sin2\beta_s^{\sss NP}$ is
$2 |r| \cos{2 \beta_s}^{\sss NP} \sin\varphi$ for all polarizations.

In summary, the $Z$FCNC model does include a contribution to the
$\bscc$ decay.  However, this effect is not large.  In addition,
because the NP operator is proportional to that of the SM, any TPs in
$\Bdecay$ are small, and no polarization-dependent indirect CP
asymmetries in $\Bsdecay$ or $\Bdecay$ are predicted.

\subsection{\boldmath $Z'$-mediated FCNC's}

In this subsection, we describe the model with $Z'$-mediated FCNC's
($Z'$FCNC).  We assume that the gauge group contains an additional
$U(1)'$, which leads to a $Z'$.  Suppose that, in the gauge basis, the
$U(1)'$ currents are \cite{Langacker:2000ju}
\beq
J_{\sss Z'}^\mu = g' \sum_i \overline{\psi}_i \gamma^\mu
\left[ \epsilon_i^{\psi_{\sss L}} P_{\sss L} +
\epsilon_i^{\psi_{\sss R}} P_{\sss R} \right] \psi_i ~,
\eeq
where $i$ is the family index, $\psi$ labels the fermions (up- or
down-type quarks, or charged or neutral leptons), and $P_{\sss L,R} =
(1 \mp \gamma_5) / 2$.  According to some string-construction or GUT
models such as $E_6$, it is possible to have family non-universal $Z'$
couplings.  That is, even though $\epsilon_i^{\sss L,R}$ are diagonal,
the couplings are not family universal.  After rotating to the
physical basis, FCNC's generally appear at tree level in both the
left-handed (LH) and right-handed (RH) sectors.  Explicitly,
\beq
B^{\psi_{\sss L}} = V_{\psi_{\sss L}} \epsilon^{\psi_{\sss
    L}} V_{\psi_{\sss L}}^\dagger ~, \qquad
B^{\psi_{\sss R}} = V_{\psi_{\sss R}} \epsilon^{\psi_{\sss
    R}} V_{\psi_{\sss R}}^\dagger ~. 
\eeq
Moreover, these couplings may contain CP-violating phases beyond
that of the SM.

In particular, $Z'{\bar b}s$ couplings can be generated:
\beq
\label{eq:zprint}
{\cal L}^{\sss Z'}_{\sss\rm FCNC} = -g' 
\left( B_{sb}^{\sss L} \, \bar s_{\sss L} \gamma_\mu b_{\sss L}
+B_{sb}^{\sss R} \, \bar s_{\sss R} \gamma_\mu b_{\sss R} \right)
\, Z^{\prime\mu} + {\rm h.c.}
\eeq
These couplings lead to a NP contribution to $\bs$-$\bsbar$ mixing
at tree level.  We define
\beq
\rho_{ff'}^{\sss L,R} \equiv \left| \frac{g' M_{\sss Z}}{g
  M_{\sss Z'}} B_{ff'}^{\sss L,R} \right| ~,
\eeq
where $g$ is the coupling of $SU(2)_{\sss L}$ in the SM.  In
Refs.~\cite{Z'FCNC,Cheung:2006tm}, it is assumed that only the LH
sector of quarks has family non-universal $U(1)'$ couplings.  Thus,
only the LH interaction above contributes to $\bs$-$\bsbar$ mixing.
It is found that one can reproduce the measured value of $\Delta M_s$
if
\beq
\rho_{sb}^{\sss L} \sim 10^{-3} ~.
\eeq
If one assumes that $g' M_{\sss Z} / g M_{\sss Z'} \sim 0.1$, then
this implies that $|B_{sb}^{\sss L}| \sim 10^{-2}$.  Note that if one
wants to include FCNC in the RH sector as well, then the constraints
on $B_{sb}^{\sss L,R}$ would be different and more uncertain because
of more free parameters in the model.

The couplings in Eq.~(\ref{eq:zprint}) will also lead to the decay $b
\to s \bar c c$ at tree level, mediated by a virtual $Z'$.  The
amplitude is
\bea
\frac{g^{\prime 2}}{M_{\sss Z'}^2} \, \left( B_{sb}^{\sss L}
\, \bar s_{\sss L} \gamma_\mu b_{\sss L} + B_{sb}^{\sss R} \,
\bar s_{\sss R} \gamma_\mu b_{\sss R} \right)
\left( B_{cc}^{\sss L} \bar c_{\sss L} \gamma^\mu c_{\sss L}
+ B_{cc}^{\sss R} \bar c_{\sss R} \gamma^\mu c_{\sss R}
\right) ~.
\eea
There are thus four types of operators, $O_{\sss LL}$, $O_{\sss LR}$,
$O_{\sss RL}$, and $O_{\sss RR}$.

However, above we have noted that the matrix element $\langle J/\psi |
\bar c \gamma^\mu \gamma_5 c | 0 \rangle$ vanishes, so that it is only
the $\gamma^\mu$ piece of these operators which contributes.  In other
words, we are left with two kinds of operators -- $O_{\sss LV}$ (as in
the SM) and $O_{\sss RV}$ -- and they can therefore be partially
combined.  (If we make the same assumption as in
Refs.~\cite{Z'FCNC,Cheung:2006tm}, then there is no $O_{\sss RV}$
operator.)  In the following, we still present the general formulae
with both LH and RH FCNC interactions.  But when computing numerical
values, we restrict ourselves to LH only.

In summary, within factorization, the total Hamiltonian can be written
as
\bea
H_{eff}^{tot} & = & H_{eff}^{\sss SM}+H_{eff}^{\sss NP,Z'}
~,\nn\\
H_{eff}^{\sss NP,Z'} & = & \frac{2\gf}{\sqrt{2}} \,
(\rho_{cc}^{\sss L} + \rho_{cc}^{\sss R}) \, \left(
\rho_{sb}^{\sss L} V_{\sss LV} + \rho_{sb}^{\sss R} V_{\sss
RV} \right) + {\rm h.c.} ~, \nn\\
V_{\sss RV} & = & \bar{s}\gamma_{\mu} ( 1+\gamma_5) b \, 
\bar{c}\gamma^{\mu} c ~,
\eea
where $H_{eff}^{\sss SM}$ and $V_{\sss LV}$ have been defined in
Eq.~(\ref{heff_final_Z}). If we suppose that $B_{sb}^{\sss R} = 0$,
then we can combine the SM and $Z'$FCNC contributions.  The ratio of
the $Z'$ contribution to the SM for the $LV$ operator only is
\bea
\left\vert \frac{2 (\rho_{cc}^{\sss L} + \rho_{cc}^{\sss R} )
\rho_{sb}^{\sss L}}{V_{cb}V_{cs}^* a_2} \right\vert & \sim &
\frac{2 \times 0.1 \, |B_{cc}^{\sss L} + B_{cc}^{\sss R}|
  \times 0.001}{0.041 \times 0.17}
= 0.03 \, |B_{cc}^{\sss L} + B_{cc}^{\sss R}| ~.
\eea
(The NP and SM matrix elements are the same, so they cancel in the
ratio.)  If $|B_{cc}^{\sss L} + B_{cc}^{\sss R}| \lsim 3$, then the NP
contribution to the $\bscc$ decay is not significant (as in the
$Z$FCNC model). But if $|B_{cc}^{\sss L} + B_{cc}^{\sss R}| \gsim 3$,
then the NP {\it does} contribute significantly to the decay, and the
analysis of $\Bsdecay$ should be modified to take this into account.
In either case, the NP contribution has the same form as that of the
SM, so that the TPs in $\Bdecay$ and the corrections leading to
polarization-dependent indirect CP asymmetries are small.

However, this can change if $B_{sb}^{\sss R} \ne 0$.  In this case,
the NP contribution is not proportional to that of the SM.  Depending
on what the flavor-changing $Z'$ couplings are, the contribution to
the decay can be significant, and there can be non-negligible
contributions to TPs and polarization-dependent indirect CP
asymmetries. (Of course, in the case of TPs, constraints from
$\Bdecay$ must be taken into account.)

\subsection{Two-Higgs-Doublet Model}

Here we examine the model with two Higgs doublets (2HDM)
\cite{2HDM}. In this model, the $\bscc$ decay occurs at tree level, as
in the SM, because of the presence of the charged Higgs boson.  The
Lagrangian for the $H^{\pm}ff^\prime$ interaction is given by
\beq
{\cal L}^{\sss 2HDM}_{H^{\pm}f_if^\prime_j} =
\frac{g}{2\sqrt{2} M_{\sss W}} H^+ \bar f_{i}
(A_{ij}+B_{ij}\gamma_5) f^\prime_{j} + {\rm h.c.},
\label{t2hdmInt}
\eeq
where $f$ and $f^\prime$ correspond to the up-type and the down-type
quarks, respectively. The couplings $A$ and $B$ depend on the set of
underlying assumptions of the particular model.  In our analysis, we
study the following two scenarios: (a) the Lagrangian for the $\bscc$
transition has a similar structure to that of the 2HDM of type II
\cite{thesis}, and (b) the Lagrangian for the $\bscc$ transition has a
similar structure to that of the so-called top-quark 2HDM
\cite{kao}. Note: the scenarios we consider are not identical to
either the 2HDM of type II or to the top-quark 2HDM, and so the
constraints on these models do not necessarily apply here. In both
cases, CP-violating phases are added to the operators in the effective
Hamiltonian.

In the most general model with new scalars, there is a flavor-changing
$H^0{\bar b}s$ coupling.  A tree-level NP contribution to
$\bs$-$\bsbar$ mixing is then produced by $H^0$ exchange. This can be
competitive with the SM and, if the coupling includes a CP-violating
phase, can lead to a significant $\bs$-$\bsbar$ mixing phase.  This
scenario occurs in little-Higgs \cite{littleHiggs} and unparticle
\cite{unparticle} models, both of which have been proposed to explain
the CDF/D0 $\Bsdecay$ data.

Of course, the exchanged particle also contributes to the decay
$\bscc$. However, here the $c{\bar c}$ quark pair forms a spin-1
$J/\psi$, which cannot be produced by the spin-0 $H^0$. Thus, the
$H^0{\bar b}s$ coupling does not lead to a contribution to the decay
$\Bsdecay$. On the other hand, if the model contains other particles,
such a contribution may be possible. For example, the unparticle model
includes flavor-changing vector particles. In this case, the size of
the contribution to the decay $\Bsdecay$ must be estimated to
determine whether the analysis of $\Bsdecay$ needs to be modified.

Now, in the 2HDM scenarios considered in this paper, there is no
flavor-changing $H^0{\bar b}s$ coupling.  As a result, the only NP
contribution to $\bs$-$\bsbar$ mixing comes from a box diagram which
involves charged-Higgs exchange.  Unless the couplings to quarks are
taken to be very large, these diagrams are smaller than the SM
contribution, so that we still have $\beta_s \simeq 0$. Therefore, the
2HDM cannot explain the CDF/D0 $\Bsdecay$ data through new effects in
$\bs$-$\bsbar$ mixing.

However, as we will see below, there can be a significant contribution
to $\bscc$ decay.  We find $|r_\lambda|$ in the range of 10-15\%,
which leads to $\sin{2 \beta}_s^{meas} =$ 0.2-0.3.  Although this
cannot account for the central values of the current data, if future
measurements find that the indirect CP asymmetry in $\Bsdecay$ is
smaller than is presently found, but is still nonzero, the 2HDM could
be the explanation. Thus, this is an example of a NP model which
affects $\Bsdecay$ through new effects in the decay.  The
$\bs$-$\bsbar$ mixing phase is still quite small, and the analysis of
$\Bsdecay$ must be redone to take this into account.

In cases (a) and (b), we have
\bea
{\rm case~(a):} ~~~ A_{ij}^{(a)} & = &
V_{ij}(m_j\tan\beta+m_i\cot\beta)\;, \nn \\
B_{ij}^{(a)} & = & V_{ij}(m_j\tan\beta-m_i\cot\beta)\;, \\
{\rm case~(b):} ~~~ A_{ij}^{(b)} & = &
V_{ij}(m_j-m_i)\tan\beta\;, \nn \\
B_{ij}^{(b)} & = & V_{ij}(m_j+m_i)\tan\beta\;,
\eea
where $\tan\beta$ is the ratio of the two vacuum expectation values of
the Higgs doublets. $V_{ij}$ is the $ij$ CKM matrix
element. Performing a Fierz transformation, we find that, within
factorization, the total effective Hamiltonian is given by
\bea 
H_{eff}& = & H_{eff}^{\sss SM}+H_{eff}^{\sss NP,2HDM}~,\nn\\
H_{eff}^{\sss NP,2HDM} & = & \frac{G_F}{\sqrt{2}} V_{cb}V_{cs}^*
\bigl[a_{\sss RR} T_{\sss RR} + a_{\sss LL}T_{\sss LL} 
+ a_{\sss RV}V_{\sss RV} +a_{\sss
LV}V_{\sss LV} \bigr] + {\rm h.c.}, \nn\\
T_{\sss RR} & = & \frac{1}{4}\bar{s}\sigma_{\mu\nu} (
1+\gamma_5) b \, \bar{c}\sigma^{\mu \nu} (1+ \gamma_5) c~, \nn\\
T_{\sss LL} & = & \frac{1}{4}\bar{s}\sigma_{\mu \nu} (
1-\gamma_5) b \, \bar{c}\sigma^{\mu\nu} (1- \gamma_5) c~, \nn\\
V_{\sss RV} & = & \bar{s}\gamma_{\mu} ( 1+\gamma_5) b
\, \bar{c}\gamma^{\mu} c~, \nn\\
V_{\sss LV} & = & \bar{s}\gamma_{\mu} ( 1-\gamma_5) b
\, \bar{c}\gamma^{\mu} c~,
\label{heff_2HDM}
 \eea 
where $H_{eff}^{\sss SM}$ is defined in Eq.~(\ref{heff_final_Z}), and
we have dropped the scalar operators which do not contribute to this
decay within factorization.

Keeping only the contributions expected to be dominant, we then find
that
\bea 
{\rm case~(a):}~~ 
H_{eff}^{\sss 2HDM(a)} & = & \frac{\gf}{\sqrt{2}} \,
V_{cb}V_{cs}^* \, a_{\sss RV} \, V_{\sss RV} ~,\nn\\
a_{\sss RV} & = & -\frac{1}{2N_c}\frac{m_bm_s}{m_{\sss
H^\pm}^2} \tan^2\beta \, e^{i \varphi^{(a)}} ,
\label{WC}
\eea 
where we have assumed a large $\tan\beta$, and 
\bea 
{\rm case~(b):}~~
H_{eff}^{\sss 2HDM(b)} & = & \frac{\gf}{\sqrt{2}} \,
V_{cb}V_{cs}^* \, a_{\sss RR} \, T_{\sss RR} ~, \nn\\
a_{\sss RR} & = & \frac{1}{2N_c}\frac{m_bm_c}{m_{\sss H^\pm}^2}
\tan^2\beta \, e^{i \varphi^{(b)}} ~.
\eea 
The new CP-violating phases $\varphi^{(a),(b)}$ have been added to the
operators in the effective Hamiltonian.  In both cases, there is one
NP operator, so these weak phases can be taken to be
polarization-independent.

In order to estimate the contributions to the indirect CP-asymmetry
and to the TP asymmetries in $\Bsdecay$, we have taken $M_{\sss B}^2
\tan^2\beta / m_{\sss H^\pm}^2 \simeq 0.5~(0.05)$ for the first
(second) case \cite{Lu-Nierste}, where $M_{\sss B}$ is the $B$-meson
mass.  With the known form factors, this allows us to calculate the
allowed values for the ratios $r_\lambda$ that we use in the following
analysis. In order to evaluate the form factors, we consider the
Melikhov-Stech \cite{Melikhov} and Ball-Zwicky \cite{Ball} models (see
Appendix for details). However, no significant difference is found
between the two models.

We begin with case (a). The first important observation is that the
ratios $r_\lambda$ depend on $\lambda$, in contrast to the $Z$FCNC and
$Z'$FCNC models. This is because $V_{\sss RV}$ (2HDM) and $V_{\sss
  LV}$ (SM) have opposite chiralities. Although their matrix elements
are the same in magnitude for each amplitude in the polarization
basis, there is a relative $-$ sign between the $\lambda=\{ 0,\| \}$
and $\lambda=\perp$ contributions.  This leads to a
polarization-dependent prediction for $\sin2\beta_s^{meas}$.  In
addition, the different chiral structures of the SM and 2HDM operators
may generate large TPs.

The expression for $\sin2\beta_s^{meas}$ is still given by
Eq.~(\ref{sin2betaZ}) (with $\sin2\beta_s^{\sss NP} = 0$). However,
there is a change in the sign assignments in front of $|r|$.  Now the
$+$ ($-$) sign applies to $\lambda = \|$ ($\lambda = 0,\perp$). (An
explicit sign in the Wilson coefficient for $a_{\sss RV}$
[Eq.~(\ref{WC})] has been included here.).  We find $|r| \simeq
0.12$. Setting $\delta_0^a \simeq \pi$, $\delta_\perp^a, \delta_\|^a
\simeq 0$, $\sin2\beta_s^{meas}$ is then $\pm 0.24 \sin\varphi^{(a)}$
for $\lambda=0,\|$ ($+$ sign) and $\lambda=\perp$ ($-$ sign).

As mentioned above, TPs are expected to be large. We compute the two
TP asymmetries, ${\cal A}_{\sss TP}^{(1,2)}$. Taking the same values
for the ratio $|r|$ and the strong phases $\delta^a_{\lambda}$ as in
the previous paragraph , we obtain $|{\cal A}_{\sss TP}^{(1)}| \lsim
0.14$ and $|{\cal A}_{\sss TP}^{(2)}| \lsim 0.07$, where the maximum
values occur at $\varphi^{(a)} = \pm \pi / 2$.

In this case, the maximum value of $|{\cal A}_{\sss TP}^{(1)}|$ is in
conflict with the data from $\Bdecay$ [Eq.~(\ref{TPresults})]. In
order to resolve this, we must take $M_{\sss B}^2 \tan^2\beta /
m_{\sss H^\pm}^2 \simeq 0.34$ in case (a). This leads to $|r| \simeq
0.08$, which in turn yields predictions of $\sin2\beta_s^{meas} = \pm
0.16 \sin\varphi^{(a)}$ (polarization-dependent indirect
CP-asymmetry), $|{\cal A}_{\sss TP}^{(1)}| \lsim 0.09$ and $|{\cal
  A}_{\sss TP}^{(2)}| \lsim 0.04$. We therefore see that, in this
case, the constraints from $\Bdecay$ reduce the effect of NP in the
decay.

Now consider case (b).  In this scenario, the ratios $r_\lambda$ also
depend on $\lambda$. The reason is that the matrix elements of the
operator $T_{\sss RR}$ entering a given polarization amplitude are
different from those of the corresponding SM contributions. This in
turn implies that there is a polarization-dependent prediction for
$\sin2\beta_s^{meas}$.  In addition, due to the different Lorentz
structures of $T_{\sss RR}$ (2HDM) and $V_{\sss LV}$ (SM), we may
expect to have large TP asymmetries.

Here, $\sin2\beta_s^{meas}$ is given by Eq.~(\ref{beta-lambda}) (with
$\sin2\beta_s = 0$). For $\delta_0^a \simeq \pi$, $\delta_\perp^a,
\delta_\|^a \simeq 0$, the NP corrections to $\sin2\beta_s^{meas}$ are
$2 |r_\lambda| \sin\varphi^{(b)}$, where $|r_\lambda| = 0.01, 0.10,
0.11$ for $\lambda=0,\|,\perp$, respectively.  We therefore see that
there is a polarization-dependent prediction for $\sin2\beta_s^{meas}$
in this scenario.

For the TP asymmetries, we find that $|{\cal A}_{\sss TP}^{(1)}| \lsim
0.06$ and $|{\cal A}_{\sss TP}^{(2)}| \lsim 0.004$, where the maximum
values occur at $\varphi^{(b)} = \pm \pi / 2$. The suppression in
${\cal A}_{\sss TP}^{(2)}$ is due to the similar sizes of $r_\|$ and
$r_\perp$, and because $\delta_\perp^a, \delta_\|^a \simeq 0$. In this
case, we have checked that the measured TP asymmetries in $\Bdecay$
[Eq.~(\ref{TPresults})] do not reduce the effects of NP in this decay.

Now, the presence of NP in $\bscc$ can produce deviations in the
measurement of $\beta$ in $\Bpdecay$. We have therefore examined
whether the constraints imposed by Eq.~(\ref{ratio}) restrict the NP
effects in $\Bsdecay$. In case (a), $r$ is simply the ratio between
the 2HDM and SM Wilson coefficients, as the relevant matrix elements
are the same in both models and thus cancel. We find $|r| \simeq
0.08$, which is allowed within a 1$\sigma$ variation. In case (b), by
following the analysis given in Refs.~\cite{HeHou,Wise,BSW}, we obtain
$|r| \simeq 6 \times 10^{-3}$. We therefore see that the present
measurement of $\beta$ in $\Bpdecay$ does not reduce the effect of the
NP in $\Bsdecay$ in either case.

In summary, we have seen that, in both cases (a) and (b) of the 2HDM,
the NP in the decay is small, but still significant, with $|r_\lambda|
= O(10\%)$. Both predict a polarization-dependent indirect CP
asymmetry in $\Bsdecay$, with $\sin2\beta_s^{meas}$ varying by 0.2-0.3
for different values of $\lambda$. Small but nonzero TP asymmetries
are also expected. ${\cal A}_{\sss TP}^{(1)}$ is found to be at most
of order (5-10)$\%$ in both scenarios, whereas ${\cal A}_{\sss
  TP}^{(2)}$ is this size only in the first.

\subsection{Supersymmetry}

We now examine the effect of supersymmetry (SUSY) on $\Bsdecay$. The
discussion is based on the analysis in Ref.~\cite{SUSY}. There it was
shown that the experimental measurements of the mass difference
$\Delta M_{\sss B_s}$ and the mercury electric dipole moment
significantly constrain the SUSY contribution to $\bs$-$\bsbar$
mixing: the total effect is just $\sin 2 \beta_s^{\sss NP} \lsim
0.1$. Thus, the only way to explain the CDF/D0 $\Bsdecay$ data is
through a SUSY contribution to the decay.  This occurs principally
through the one-loop correction to $\btos$ from gluino exchange.

Here we extend the previous analysis in several ways. The focus of
Ref.~\cite{SUSY} was mainly on the transverse amplitudes, as the
predictions for the indirect CP asymmetries were found to be
independent of hadronic form factors. (For completeness, we repeat
these results below.) In this paper we also calculate the indirect CP
asymmetry for the longitudinal amplitude, which does depend on the
hadronic form factors. (However, the effect of SUSY on the
longitudinal amplitude is found to be small). We also compute the TP
asymmetries in $\Bdecay$ and $\Bsdecay$ in the presence of SUSY, as
well as its contribution to $\Bpdecay$.

There are two main operators that can contribute to the transition
$\bscc$, both of dipole type.  After a Fierz transformation (and
neglecting the color-octet pieces), they are
\bea
H_{\sss SUSY} & = & C_g O_g + \widetilde{C_g} \widetilde{O_g}
\nn\\
O_g & = & Y_g \left[ -\frac{2}{N_c} \left(\bar{s}_{\alpha}
\gamma_{\mu} \frac{{\sla{q}}}{m_b} (1+\gamma_5)
b_{\alpha}\right) \left( c_{\beta} \gamma^{\mu}
c_{\beta}\right) \right],\nn\\
\widetilde{O}_g & = & Y_g \left[ -\frac{2}{N_c} \left(
\bar{s}_{\alpha} \gamma_{\mu} \frac{{\sla{q}}}{m_b}
(1-\gamma_5) b_{\alpha}\right) \left( c_{\beta}
\gamma^{\mu} c_{\beta} \right) \right].
\label{susy_np}
\eea
Here, $Y_g = -(\alpha_s m_b^2)/(4\pi m_{\sss J/\psi}^2)$; $q$ is the
momentum transfer.  The coefficients $C_g$ and $ \widetilde{C_g}$ can
be found in Ref.~\cite{SUSY} and depend on the mass insertions $ \lr$
and $\rl$.

With the operators above, we can compute the ratio of the SUSY and SM
contributions for the transverse polarizations \cite{SUSY}:
\beq
|r_{\|}| e^{i\varphi^\|_{{\sss SUSY}}} = \frac{(Y- \widetilde{Y})}{X} ~~,~~~~
|r_{\perp}| e^{i\varphi^\perp_{{\sss SUSY}}} = \frac{(Y+
 \widetilde{Y})}{X} ~,
\label{ratio_susy}
\eeq
where
\beq
Y = \frac{\sqrt{2}C_g}{\gf}Y_g \left[ -\frac{2}{N_c}\right]
~,~~ \widetilde{Y} = \frac{\sqrt{2}\widetilde{C_g}}{\gf}Y_g
\left[ -\frac{2}{N_c}\right] ~, ~~ 
X=V_{cb}V_{cs}^* (c_2 + { c_1 \over N_c}) \approx 0.007. \
\eeq
As is clear from these expressions, the form factors and other
hadronic quantities cancel in the ratios. We therefore obtain clean
predictions for $r_{\|}$ and $r_{\perp}$ in SUSY.

We take the following values for the masses: $m_{\tilde g}=m_{\tilde
  q}=500$ GeV and $m_b(m_b)=4.5$ GeV, obtaining
\bea
Y & \approx & 2.13 \, (\delta^d_{\sss LR})_{23}
\left[\frac{-2}{N_c}Y_g\right] = 0.05 \, (\delta^d_{\sss
  LR})_{23} ~,\nn\\
\widetilde{Y} & \approx & 2.13 \, (\delta^d_{\sss RL})_{23}
\left[\frac{-2}{N_c}Y_g\right] =0.05 \, (\delta^d_{\sss
RL})_{23} ~.
\label{ynum}
\eea
Eq.~(\ref{ratio_susy}) then gives
\bea
\label{mag}
\left \vert r_{\|} \right \vert & \approx & 7 \Big[ \vert\lr\vert^2 + \vert
  \rl \vert^2
-~2 \vert\lr\vert\vert \rl \vert \cos (\varphi_{\sss
LR}-\varphi_{\sss RL}) \Big]^{1/2} ~, \nn\\
\left \vert r_{\perp} \right \vert & \approx & 7 \Big[ \vert\lr\vert^2 + \vert
  \rl \vert^2
+~2 \vert\lr\vert\vert \rl \vert \cos (\varphi_{\sss
LR}-\varphi_{\sss RL}) \Big]^{1/2} ~,
\eea
where $\varphi_{\sss LR}$ and $\varphi_{\sss RL}$ are the phases of
$\lr$ and $\rl$, respectively. The phases of the helicity amplitudes
can be obtained from
\bea
\label{phase}
\vert r_\| \vert \sin \varphi^\|_{{\sss SUSY}} & \approx &
7~\left[\vert\lr\vert \sin{\varphi_{\sss
LR}}-\vert \rl \vert \sin{\varphi_{\sss RL}}\right] ~, \nn\\
\vert r_{\perp} \vert \sin \varphi^{\perp}_{{\sss SUSY}} & \approx & 
 7~\left[\vert\lr\vert \sin{\varphi_{\sss
LR}}+\vert \rl \vert \sin{\varphi_{\sss RL}} \right] ~.
\eea

In order to calculate the quantities $|r_{\|}|$ and $|r_{\perp}|$, we
need values for the square roots of Eq.~(\ref{mag}). Below, we
consider several different realistic scenarios for the magnitudes and
phases of $\lr$ and $\rl$. In all cases, each of the square roots
takes a value between 0 and 0.02, so that $|r_{\|}|$ and $|r_{\perp}|$
can be (independently) in the range 0-14\%. We therefore see that
these ratios can be significant ($\gsim 10\%$) within SUSY.

We now turn to the longitudinal amplitude. In this case, the ratio
$|r_{0}|$ will depend on form factors. The SUSY amplitude for the
longitudinal polarization is
\bea
\label{A0defsusy2}
A^0_{\sss SUSY}&=& \frac{\gf}{\sqrt{2}}  m_{\sss J/\psi}g_{\sss J/\psi} 
(Y-\tilde{Y}) \Big[ \xi_1  + \xi_2 \Big],
\eea
where
\bea
\label{xidef1}
\xi_1&=& \Big[ (m_{\sss B_s}+m_{\phi}) A_1(m^2_{\sss
J/\psi})x - \frac{2m_{\sss J/\psi} m_{\phi}}{(m_{\sss B_s}+m_{\phi})}
 A_2(m^2_{\sss J/\psi}) (x^2-1) \Big], \\ \nn
\xi_2 &=& \frac{2m_{\sss J/\psi} m_{\phi}}{m^2_b} (x^2-1) \nn \\
&& \times \Bigg[\Big[-
(m_{\sss B_s}+m_{\phi}) A_1(m^2_{\sss J/\psi}) 
+ \frac{A_2(m^2_{\sss J/\psi})}
{{m_{\sss B_s}+m_{\phi}}} \Big( m^2_{\sss B_s} +
 ( m^2_{\sss B_s}+ m^2_{\phi} - m^2_{\sss J/\psi})/2 \Big)\Big] \nn \\
&&
+~\frac{( m^2_{\sss B_s}+ m^2_{\sss J/\psi}- m^2_{\phi}) }{2 m^2_{\sss J/\psi}}
\Big[(m_{\sss B_s}+m_{\phi})
A_1(m^2_{\sss J/\psi})- A_2(m^2_{\sss J/\psi})
(m_{\sss B_s}-m_{\phi}) \nn \\
&&
-~2  m_{\phi} A_0(m^2_{\sss J/\psi}) \Big] \Bigg]  ~.
\eea
The various hadronic form factors in the expression above are defined
in the Appendix.
We find that $|r_{0}|$ is given by
\beq
\label{ratiolongSMSUSYdef1}
|r_{0}| = \left| \frac{A^0_{\sss SUSY}}{A^0_{\sss SM}} \right| = \Big[1+
\frac{\xi_2}{\xi_1} \Big] \left| \frac{ (Y-\tilde{Y})}{X}
\right| ~.
\eeq

In order to compute $|r_0|$, we have to use a model to calculate the
form factors. We consider the models by Melikhov-Stech (MS)
\cite{Melikhov} and Ball-Zwicky (BZ) \cite{Ball}. Using the results of
the form factors (see Appendix for details) one can obtain predictions
for $|r_0|$:
\bea
|r_0| & \approx & C \times \Big[ \vert\lr\vert^2 + \vert
  \rl \vert^2
- 2 \vert\lr\vert\vert \rl \vert \cos (\varphi_{\sss
LR}-\varphi_{\sss RL}) \Big]^{1/2} ~,~~
\eea
where $C=0.8$ (MS) or 0.9 (BZ). Since, as noted above, the square root
takes a maximum value of 0.02, $|r_0|$ is quite small, due to a
cancellation between the various NP amplitudes. We therefore see that,
although the SUSY contribution to the transverse-amplitude ratios
$|r_{\|}|$ and $|r_{\perp}|$ can be significant, the contribution to
the longitudinal-amplitude ratio $|r_0|$ is negligible.

We can now estimate the contributions to the indirect CP asymmetry in
$\Bsdecay$ and to the TP asymmetries. Since one has different
contributions to the three $r_\lambda$'s in SUSY, these effects may be
important. We consider the following four scenarios for the magnitudes
and phases of $\lr$ and $\rl$:
\begin{enumerate}

\item $\vert\lr\vert = \vert\rl\vert = 0.01$ and
$(\varphi_{\sss LR}-\varphi_{\sss RL}) = 0$,

\item $\vert\lr\vert = \vert\rl\vert = 0.01$ and
$(\varphi_{\sss RL}-\varphi_{\sss LR}) = \pi$,

\item $\vert\lr\vert = 0.01$, $\rl = 0$,

\item $\lr = 0$, $\vert\rl\vert = 0.01$.

\end{enumerate}
(The value of 0.01 for $\vert\lr\vert$ and/or $\vert\rl\vert$ is
consistent with the constraint from $b \to s \gamma$ \cite{SUSY}.)

For the indirect CP asymmetry in $\Bsdecay$, one obtains the following
result:
\bea
\left. \sin{2 \beta}_s^{meas} \right\vert_0 & = & \sin 2
\beta_s^{\sss NP} ~, \nn\\
\left. \sin{2 \beta}_s^{meas} \right\vert_\| & = & 
\sin 2 \beta_s^{\sss NP} + 2 \vert r_\|
\vert \cos 2 \beta_s^{\sss NP} \sin \varphi^\|_{{\sss SUSY}}
\cos \delta_\|^a ~, \nn\\
\left. \sin{2 \beta}_s^{meas} \right\vert_\perp & = &
\sin 2 \beta_s^{\sss NP} + 2 \vert r_\perp
\vert \cos 2 \beta_s^{\sss NP} \sin \varphi^\perp_{{\sss
SUSY}} \cos \delta_\perp^a ~.
\eea
In the above, we have set the SM $\beta_s = 0$. $\beta_s^{\sss NP}$ is
the SUSY contribution to $\bs$-$\bsbar$ mixing,
$\varphi^{\|,\perp}_{{\sss SUSY}}$ are the SUSY weak phases of the
$\|$ and $\perp$ amplitudes, and $\delta_{\|,\perp}^a$ are the SM
strong phases (both $\approx 0$). The values of $\vert r_\| \vert$,
$\vert r_\perp \vert$, $\varphi^{\|}_{{\sss SUSY}}$ and
$\varphi^{\perp}_{{\sss SUSY}}$ are given in Table~\ref{rTtable} using
Eqs.~(\ref{mag}) and (\ref{phase}). We see that SUSY does indeed
predict a polarization-dependent indirect CP asymmetry in $\Bsdecay$,
with the value of $\sin{2 \beta}_s^{meas}$ in different polarizations
varying by as much as 0.28. Note: the effect is largest in scenarios
(1) and (2). However, even in scenarios (3) and (4), for which the
contributions to $\vert r_\| \vert$ and $\vert r_\perp \vert$ are not
considered significant, one has a maximal difference of $0.14$ for
$\sin{2 \beta}_s^{meas}$ in the longitudinal and transverse
polarizations. This may be measurable.

\TABLE{
\begin{tabular}{lllll}
\hline
scenario & \hspace{0.5cm} $\vert r_\| \vert$ & \hspace{0.5cm}
$\vert r_\perp \vert$ &\hspace{0.5cm} $\varphi^{\|}_{{\sss
SUSY}}$ & \hspace{0.5cm}$\varphi^{\perp}_{{\sss SUSY}}$ \\
\hline
(1) & \hspace{0.5cm} 0 & \hspace{0.5cm} 0.14 & \hspace{0.5cm}
0 & \hspace{0.5cm} $\varphi_{\sss LR}$\\
(2) & \hspace{0.5cm} 0.14 &\hspace{0.5cm} 0 & \hspace{0.5cm}
 $\varphi_{\sss LR}$& \hspace{0.5cm} 0 \\
(3) & \hspace{0.5cm} 0.07 & \hspace{0.5cm} 0.07 &
 \hspace{0.5cm} $\varphi_{\sss LR}$ &\hspace{0.5cm} $\varphi_{\sss LR}$
 \\
(4) & \hspace{0.5cm} 0.07 & \hspace{0.5cm} 0.07 &
\hspace{0.5cm} $-\varphi_{\sss RL}$ & \hspace{0.5cm}
$\varphi_{\sss RL}$ \\ 
\hline
\end{tabular}
\caption{SUSY predictions for $\vert r_\| \vert$, $\vert r_\perp
  \vert$, $\varphi^{\|}_{{\sss SUSY}}$ and $\varphi^{\perp}_{{\sss
      SUSY}}$ in the four scenarios for the magnitudes and phases of
  $\lr$ and $\rl$ described in the text.}
\label{rTtable}
}

We now move to the predictions of triple-product asymmetries in
$\Bsdecay$.  Using the experimental strong phases $\delta^a_0 \approx
\pi$, $\delta^a_\perp, \delta^a_\parallel \approx 0$ and the values of
the form factors given in the Appendix, the TP asymmetries can be
calculated. They are shown in Table~\ref{TPtable} (there is no
difference between the MS and BZ predictions). We see that effects of
up to 5-10\% are allowed. (We have explicitly calculated the TP
asymmetries in $\Bdecay$, and find that they are equal to those in
$\Bsdecay$. There is therefore no conflict with the $\Bdecay$ data.)

\TABLE{
\begin{tabular}{lll}
\hline
scenario & \hspace{0.5cm} ${\cal A}_{\sss TP}^{(1)}$ &
\hspace{0.5cm} ${\cal A}_{\sss TP}^{(2)}$ \\
\hline
(1) & \hspace{0.5cm} $\approx 0.09 \sin{\varphi_{\sss LR}}$ &
\hspace{0.5cm} $\approx 0.08 \sin{\varphi_{\sss LR}}$ \\
(2) & \hspace{0.5cm} $ 0 $ & \hspace{0.5cm} $\approx -0.08
\sin{\varphi_{\sss LR}}$ \\
(3) & \hspace{0.5cm} $\approx 0.04 \sin{\varphi_{\sss LR}}$
& \hspace{0.5cm} 0 \\
(4) & \hspace{0.5cm} $\approx 0.04 \sin{\varphi_{\sss RL}}$
& \hspace{0.5cm} $\approx 0.08 \sin{\varphi_{\sss RL}}$ \\
\hline
\end{tabular}
\caption{SUSY predictions for the triple-product asymmetries ${\cal
    A}_{\sss TP}^{(1)}$ and ${\cal A}_{\sss TP}^{(2)}$ in the four
  scenarios for the magnitudes and phases of $\lr$ and $\rl$ described
  in the text.}
\label{TPtable}
}

Now, if there is new physics in the decay $\bscc$, it can affect the
measurement of $\beta$ in $\Bpdecay$. Earlier we said that a NP
contribution of $\le 15\%$ is still permitted. However, in the case of
SUSY, we can do an explicit calculation. We find that the ratio of
SUSY and SM amplitudes in $\Bpdecay$ is
\beq
\label{rAJpsiK}
|r| = \left| \frac{A_{\sss SUSY}}{A_{\sss SM}} \right| = \left|
\frac{ (Y+\tilde{Y})}{X} \right| \left(1- \frac{\xi^{\sss
BK} (m^2_{\sss
    J/\psi})}{m^2_b \, F^{\sss BK}_1(m^2_{\sss
J/\psi}) } \right),
\eeq
where 
\begin{eqnarray}
\label{xiKdef}
\xi^{\sss BK}(m^2_{\sss J/\psi})&=& F_1(m^2_{\sss J/\psi})
\Big( m^2_{\sss B} +  \frac{m^2_{\sss B}+m^2_{\sss
    K}-m^2_{\sss J/\psi}}{2}\Big) \\ 
&&
-\Big(\frac{m^2_{\sss B} - m^2_{\sss K}}{m^2_{\sss J/\psi}} \Big)
 \Big(\frac{m^2_{\sss B}+ m^2_{\sss J/\psi}-m^2_{\sss K}}{2}\Big)
\Big(F_1(m^2_{\sss J/\psi}) 
-F_0(m^2_{\sss J/\psi})\Big),   \nn
\end{eqnarray}
and the form factors $F_{0,1}$ are defined in the Appendix.
Using $m_b(m_b)$= 4.5 GeV and the form-factor values \cite{cheng2}
\begin{eqnarray}
\label{FormFBKcheng}
F^{\sss BK}_1(m^2_{\sss J/\psi})=&  0.70~, ~~~F^{\sss BK}_0
(m^2_{\sss J/\psi})= 0.50 ~,
\end{eqnarray}
one finds
\bea
|r| & = & -0.5 \Big[ \vert\lr\vert^2 + \vert
  \rl \vert^2 
+~2 \vert\lr\vert\vert \rl \vert \cos (\varphi_{\sss
LR}-\varphi_{\sss RL}) \Big]^{1/2} .~
\eea
The square root takes a maximum value of 0.02, so that $|r|$ is quite
small. We therefore see that the SUSY contribution to $\Bpdecay$ is
suppressed due to a cancellation of the various contributing
amplitudes.

\section{Conclusions \label{sec4}}

The CDF and D0 collaborations recently made a measurement of indirect
CP violation in $\Bsdecay$ and found a $2.2\sigma$ deviation from the
standard model (SM). This suggests a nonzero value of $\beta_s$, the
phase of $\bs$-$\bsbar$ mixing. Since the SM predicts $\beta_s\simeq
0$, we assume the CDF/D0 result may be due to new physics (NP). In
this paper, we have argued that any analysis of NP in $\bs$-$\bsbar$
mixing is incomplete if NP in the decay $\bscc$ is not considered.

In fact, most models that produce new effects in $\bs$-$\bsbar$ mixing
also contribute to $\bscc$.  We have analyzed a number of such models
and find that, indeed, there can be NP effects in the decay.  In
general, the effect is not enormous.  However, it may not be
insignificant either -- we find that the ratio of NP to SM
contributions can be as large as $O$(10-15\%).  If this ratio is big
in a given model, then even if the NP contribution to $\bs$-$\bsbar$
mixing is not large enough to reproduce the CDF/D0 measurement, the
addition of the new effects in $\bscc$ may be sufficient.  Similarly,
there are certain models which do not contribute significantly to
$\beta_s$, and hence cannot account for the current data. However, if
future measurements find a smaller (nonzero) value for the indirect CP
asymmetry in $\Bsdecay$, these models might be able to explain the
data through NP contributions to the decay.

Specifically, we have examined four NP models. (In all cases,
constraints from $\Bpdecay$ have been taken into account.)  We find
that the model with $Z$-mediated FCNC's does not lead to big effects
in the $\bscc$ decay.  On the other hand, the model with $Z'$-mediated
FCNC's may do so if certain of the $Z'$ couplings are sufficiently
large.  The two-Higgs-doublet model (2HDM) contributes very little to
$\bs$-$\bsbar$ mixing.  However, it can give significant contributions
to the decay, so that the 2HDM can account for somewhat smaller values
of the indirect CP asymmetry in $\Bsdecay$.  Supersymmetry is similar
-- the contribution to $\beta_s$ is small (though nonzero), but it can
give large contributions to the decay.

The models which contribute significantly to the decay $\bscc$
typically also have other effects.  In general, they predict
polarization-dependent indirect CP asymmetries in both $\Bsdecay$ and
$\Bdecay$.  And they also predict small, but nonzero, triple-product
(TP) asymmetries in $\Bsdecay$ ($\le 10\%$), consistent with the
constraints from TP asymmetries in $\Bdecay$.

\bigskip
\noindent
{\bf Acknowledgments}:
We would like to thank M. Imbeault and J. Rosner for helpful
conversations, and A. Lenz and R. Mohanta for useful
communications. This work was financially supported by NSERC of Canada
(D.~L., M.~N., A.~S.). The work of C.~C. is supported in part by the
National Science Council of Taiwan, R.~O.~C.\ under Grant
No.~NSC~97-2112-M-008-002-MY3 and NCTS.


\appendix

\section{Form Factors and Matrix Elements}

For a general effective four-quark operator ${\cal O} \sim X\otimes Y$
($X,Y=\bar q \gamma_\mu q, \, \bar q \gamma_\mu \gamma_5 q, \, \bar q
\sigma_{\mu \nu} q$ or $\bar q \sigma_{\mu \nu}\gamma_5 q$), the
matrix element is factorized as
\beq
\bra{J/\psi \phi} {\cal O} \ket{\bs} \to \bra{\phi} X \ket{\bs} \bra{J/\psi} Y \ket{0}~,
\eeq
where $\bra{\phi} X \ket{\bs}$ is calculable using known form factors
and $\bra{J/\psi} Y \ket{0}$ is proportional to the $J/\psi$ decay constant.

For the $\bs \to \phi$ form factors, we follow the definitions in
Ref.~\cite{HeHou}:
\bea
\bra{\phi(p,\epsilon)} \bar s \gamma_\mu (1 \pm \gamma_5) b \ket{\bs(p_{\sss B_s})}
& = &
\pm i \epsilon_\mu^* (m_{\sss B_s}+m_{\sss \phi}) A_1(s) \nn\\
&& \mp~i (p_{\sss B_s}+p)_\mu (\epsilon^* \cdot p_{\sss B_s})
\frac{A_2(s)}{m_{\sss B_s}+m_{\sss \phi}}\nn\\
&& \mp i q_\mu (\epsilon^* \cdot p_{\sss
B_s}) \frac{2 m_{\sss \phi}}{s} (A_3(s)-A_0(s)) \nn\\
&& +~\epsilon_{\mu
\nu \rho \sigma} \epsilon^{*\nu}p_{\sss B_s}^\rho p^\sigma
\frac{2 V(s)}{m_{\sss B_s}+m_{\sss \phi}} ~, \nn\\
\bra{\phi(p, \epsilon)} \bar s \sigma_{\mu \nu} b \ket{\bs(p_{\sss
B_s})} 
& = & -i \epsilon_{\mu \nu \rho \sigma}\bigl[
g_+(s)\epsilon^{*\rho}(p_{\sss B_s}+p)^\sigma 
+g_-(s)\epsilon^{*\rho}q^\sigma  \nn\\
&& 
+h(s)\frac{\epsilon^* \cdot p_{\sss B_s}}{m_{\sss B_s}^2-m_{\sss \phi}^2}
(p_{\sss B_s}+p)^\rho q^\sigma
\bigr]\;,\nn\\
\bra{\phi(p, \epsilon)} \bar s \sigma_{\mu \nu}\gamma_5 b \ket{\bs(p_{\sss
B_s})} 
& = & g_+(s)[\epsilon^{*}_\mu(p_{\sss B_s}+p)_\nu
- \epsilon^{*}_\nu(p_{\sss B_s}+p)_\mu] \nn\\
&& 
+ g_-(s)[\epsilon^{*}_\mu q_\nu
- \epsilon^{*}_\nu q_\mu] \nn\\
&& \hskip-2truecm
+h(s)\frac{\epsilon^* \cdot p_{\sss B_s}}{m_{\sss B_s}^2-m_{\sss \phi}^2}
[(p_{\sss B_s}+p)_{\mu}q_{\nu}
-(p_{\sss B_s}+p)_{\nu}q_{\mu}]\;,
\eea
where $q=p_{\sss B_s}-p$ and $s=q^2$.  We consider two scenarios for
the form factors -- the Melikhov-Stech \cite{Melikhov} and Ball-Zwicky
\cite{Ball} models. The results are shown in Table \ref{tab:FFVT}.

\TABLE{
\caption{Predictions for vector ($A_1$, $A_2$, $A_0$, $V$) and tensor
  ($g_+$, $g_-$, $h$) form factors in the Melikhov-Stech and
  Ball-Zwicky models, all evaluated at $s=m_{\sss J/\psi}^2$.  The
  tensor form factors are computed in the heavy-quark effective theory
  at maximum recoil \cite{HeHou}.}
\label{tab:FFVT}
\begin{tabular}{cccccccc}
\hline
FF & $A_1$ & $A_2$ & $A_0$ & $V$ & $g_+$ & $g_-$ & $h$
\\
\hline
MS & $0.42$ & $0.49$ & $0.76$ & $0.80$
& $0.69$ & $-0.66$ & $0.18$
\\
BZ & $0.42$ & $0.38$ & $0.89$ & $0.82$
& $0.70$ & $-0.68$ & $0.30$
\\
\hline
\end{tabular}
}

We define the $J/\psi$ decay constants according to Ref.~\cite{Ball}:
\bea
\bra{J/\psi (q, \epsilon)}\bar c \gamma^\mu c \ket{0} &=& f_{\sss J/\psi} m_{\sss J/\psi}
\epsilon^{*\mu}~,\nn\\
\bra{J/\psi (q, \epsilon)} \bar c \sigma^{\mu \nu} c \ket{0}&=& -i
f_{\sss J/\psi}^\bot (\epsilon^{*\mu} q^\nu - \epsilon^{*\nu}
q^\mu)~,
\eea
which implies
\beq
\bra{J/\psi (q, \epsilon)} \bar c \sigma^{\mu \nu} \gamma_5 c \ket{0} =
-\frac{1}{2}f_{\sss J/\psi}^\bot \epsilon^{\mu \nu \rho \sigma}
(\epsilon^*_\rho q_\sigma - \epsilon^*_\sigma q_\rho)~.
\eeq
We take $f_{\sss J/\psi}^\bot \sim f_{\sss J/\psi} = 405$ MeV \cite{cheng}.

The form factors relevant for $\Bpdecay$ are defined as \cite{BSW}
\begin{eqnarray}
 \label{PsudoFF}
\langle K (k_2) |\bar{s} \gamma_\mu (1-\gamma_5) b
|B(p)\rangle
&=& R_{\mu}F_1^{\sss BK}(r^2)+Q_{\mu}F_0^{\sss BK}(r^2) ~, \nn\\
R_{\mu} &=& \Big[p_{\mu} +k_{2\mu} -\frac{m^2_{\sss B} -
m^2_{\sss K}}{r^2} r_{\mu} \Big] ~, \nn\\
Q_{\mu} &=& \frac{m^2_{\sss B}- m^2_{\sss K}}{r^2}r_{\mu} ~.
\end{eqnarray}

In order to calculate the matrix elements, we define in the $B$-meson
rest frame the four-momenta
\bea
p_{\sss B_s} & = & (m_{\sss B_s},0,0,0)~, \nn\\
p_{\sss \phi} & = & (E_{\sss \phi},0,0,-p_c)~, \nn\\
p_{\sss J/\psi} & = & (E_{\sss J/\psi},0,0,p_c)~,
\eea
and polarization four-vectors
\bea
\epsilon^0_{\sss \phi} & = & \frac{1}{m_{\sss \phi}} (p_c,
0,0,-E_{\sss \phi})~, \nn\\
\epsilon^\pm_{\sss \phi} & = & \frac{1}{\sqrt{2}}(0,\mp 1, +i, 0)~,\nn\\
\epsilon^0_{\sss J/\psi} & = & \frac{1}{m_{\sss J/\psi}} (p_c,
0,0,E_{\sss J/\psi})~, \nn\\
\epsilon^\pm_{\sss J/\psi} & = &
\frac{1}{\sqrt{2}}(0,\mp 1, -i, 0)~,
\eea
where $p_c = {[(m^2_{\sss B_s}-{(m_{\sss J/\Psi}+m_{\sss
        \phi})}^2)(m^2_{\sss B_s}-{(m_{\sss J/\Psi}-m_{\sss
        \phi})}^2)]}^{\frac{1}{2}} / 2 \, m_{\sss B_s}$.

The resulting matrix elements for a given helicity state ($\lambda= 0,
\pm$) are
\bea
\left. \bra{\phi} \bar s \sigma_{\mu \nu} b \ket{\bs} \bra{J/\psi} \bar c \sigma^{\mu \nu} c \ket{0} \right|_{\lambda=\pm} & = & \left. \bra{\phi}
\bar s \sigma_{\mu \nu} \gamma_5 b \ket{B_s} \bra{J/\psi} \bar c \sigma^{\mu \nu} \gamma_5 c \ket{0} \right|_{\lambda=\pm} \nn\\
& = &
\mp 4 i f_{\sss J/\psi}^\bot g_+(m_{\sss J/\psi}^2) 
m_{\sss B_s} p_c~,\nn\\
\left.\bra{\phi} \bar s \sigma_{\mu \nu} b \ket{\bs} \bra{J/\psi} \bar c \sigma^{\mu \nu} \gamma_5 c \ket{0} \right|_{\lambda=0} & = & \left. \bra{\phi}
\bar s \sigma_{\mu \nu} \gamma_5 b \ket{B_s} \bra{J/\psi} \bar c \sigma^{\mu \nu} c
\ket{0} \right|_{\lambda=0} \nn\\
& & \hskip-2truecm  = -~4 p_c^2 m_{\sss B_s}^2
\left[ g_+(m_{\sss J/\psi}^2)-h(m_{\sss J/\psi}^2)
m_{\sss J/\psi}^2/(m_{\sss
B_s}^2-m_{\sss \phi}^2) \right] \nn\\
& & \hskip-5truecm -~i
\frac{f_{\sss J/\psi}^\bot}{m_{\sss J/\psi} m_{\sss \phi}} \bigl[
g_+(m_{\sss J/\psi}^2)
(m_{\sss B_s}^2-m_{\sss J/\psi}^2-m_{\sss \phi}^2) 
(m_{\sss B_s}^2-m_{\sss \phi}^2+
\frac{g_-(m_{\sss J/\psi}^2)}{g_+(m_{\sss J/\psi}^2)}m_{\sss J/\psi}^2) \bigr] ~,
\nn\\
\left.\bra{\phi} \bar s \sigma_{\mu \nu} b \ket{\bs} \bra{J/\psi} \bar c \sigma^{\mu \nu} \gamma_5 c \ket{0} \right|_{\lambda=\pm} & = & \bra{\phi}\left|
\bar s \sigma_{\mu \nu} \gamma_5 b \ket{\bs} \bra{J/\psi} \bar c \sigma^{\mu \nu} c
\ket{0}\right|_{\lambda=\pm} \nn\\
&= & 2 i f_{\sss J/\psi}^\bot
g_+(m_{\sss J/\psi}^2) 
\bigl[m_{\sss B_s}^2-m_{\sss \phi}^2
+\frac{g_-(m_{\sss J/\psi}^2)}{g_+(m_{\sss J/\psi}^2)}m_{\sss J/\psi}^2\bigr] ~,
\nn\\
\left.\bra{\phi} \bar s \gamma_\mu b \ket{\bs} \bra{J/\psi} \bar c \gamma^\mu c \ket{0}
\right|_{\lambda=\pm} &= & \mp 2 i f_{\sss J/\psi} V \frac{m_{\sss J/\psi}
m_{\sss B_s} p_c}{m_{\sss B_s}+m_{\sss \phi}} ~,\nn\\
\left.\bra{\phi}  \bar s \gamma_\mu \gamma_5 b \ket{\bs} \bra{J/\psi} \bar c \gamma^\mu c \ket{0}
\right|_{\lambda=0} &=& i f_{\sss J/\psi} \frac{m_{\sss B_s}+m_{\sss
\phi}}{2 m_{\sss \phi}} \times \nn\\
&& 
\left\{ (m_{\sss B_s}^2-m_{\sss J/\psi}^2-m_{\sss
\phi}^2)A_1 
 -~\frac{4 m_{\sss B_s}^2 p_c^2
A_2 }{(m_{\sss B_s}+m_{\sss \phi})^2} \right\}~,\nn\\
\left.\bra{\phi}  \bar s \gamma_\mu \gamma_5 b \ket{\bs} \bra{J/\psi} \bar c \gamma^\mu \ket{0}
\right|_{\lambda=\pm}&=& -i f_{\sss J/\psi} m_{\sss J/\psi} (m_{\sss B_s}+m_{\sss
\phi})A_1~.
\eea  
The remaining matrix elements are zero.


\end{document}